# Surface-Induced Phase Transition During Coalescence of Au Nanoparticles: A Molecular Dynamics Simulation Study[1]

Reza Darvishi Kamachali

Contact Address: Max-Planck-Institut für Eisenforschung, Max-Planck-Str. 1, 40237 Düsseldorf, Germany. Email: kamachali@mpie.de

**Abstract**

In this study, the melting and coalescence of Au nanoparticles were investigated using molecular dynamics simulation. The melting points of nanoparticles were calculated by studying the potential energy and Lindemann indices as a function of temperature. The simulations show that coalescence of two Au nanoparticles of the same size occurs at far lower temperatures than their corresponding melting temperature. For smaller nanoparticles, the difference between melting and coalescence temperature increases. Detailed analyses of the Lindemann indices and potential energy distribution across the nanoparticles show that the surface melting in nanoparticles begins at several hundred degrees below the melting point. This suggests that the coalescence is governed by the surface diffusion and surface melting. Furthermore, the surface reduction during the coalescence accelerates its kinetics. It is found that for small enough particles and/or at elevated temperatures, the heat released due to the surface reduction results in a melting transition of the two attached nanoparticles.

Keywords: Molecular Dynamics Simulation; Nanosystems; Coalescence; Surface Melting; Gold Nanoparticles; Phase Transition.

## 1. Introduction

Due to the unique electronic, optical and catalytic properties of metallic nanoparticles [1-3], an increasingly growing interest is developing to explore nanoparticles technological applications [4-6]. In this context, the interactions between nanoparticles and their surrounding objects such as solvents, ion/electron beams, and other nanoparticles are fundamentally and technologically important for their storage and performance [7, 8]. In particular, the strong attraction among nanoparticles due to their large surface/volume ratio makes them to stick and agglomerate that change some of their desired functional properties. Metallic/superconductor conjunctions [9], light scattering and absorption [10], and sorting [11] of nanoparticles are some applicable experimental examples which confront this problem. Hence, studying these interactions can be useful to control and improve synthesize and assembly of nanoscale materials and components.

---

[1] Completed in April 2008. See Acknowledgments for details.

The interaction among the nanoparticles is driven by a reduction in the surface energy. This can be either preserving the shape of nanoparticles at lower temperatures or changing their shape and size by their coalescence at elevated temperatures. In fact, the surface energy of a nanoparticle can be a large portion of its total energy. In this study, we explore the role of surface reduction during the coalescence of Gold (Au) nanoparticles their evolution using Molecular Dynamics (MD) Simulation method.

Au nanoparticles are one of the well-known noble metal nanoparticles that have found some important applications in biotechnology [12], in synthesis [13] as catalyst and in nanoparticles-networked film [14] as conductor and characterization devices [15] as probes. Here we conduct a systematic study of melting and coalescence behavior of Au nanoparticles. In the first step, the melting behavior of single Au nanoparticles are studies as a function of their size. Then we study the coalescence of the nanoparticles of the same size at different temperatures. Detailed analyses of Lindemann index (LI) and potential energy distribution across the nanoparticles were performed to study the melting and coalescence mechanisms of nanoparticles. The role of surface melting and surface area reduction are discussed in details, both of which are specific features of the nanosized systems.

## 2. Simulation method

In the current MD simulations, a glue potential for Au calculated by force matching method was used [16]. The force matching method is a very effective tool to obtain realistic classical potentials [17] that is based on fitting the potential to ab-initio atomic forces of many atomic configurations. A glue potential

$$V = \frac{1}{2}\sum_{ij}\phi(r_{ij}) + \sum_{i}U\left(\sum_{j}\rho(r_{ij})\right) \quad (1)$$

is defined by standard pair potential $\phi(r_{ij})$, a glue function $U(n)$ (an energy associated with coordination n of atoms), an atomic density function $\rho(r_{ij})$ (short-ranged monotonically decreasing function of distance) where $r_{ij}$ is the distance between two atoms. This potential has been successfully employed in several previous works [18, 19].

In this study, the constant temperature simulations were carried out on Au nanoparticles with 586, 1289 and 2406 atoms. All nanoparticles were initially constructed to a prefect truncated polyhedral structure (Figure 1a). The Verlet velocity algorithm was employed to solve motion equations and, the desired temperature was obtained by uniform kinetic energy scaling method [20]. The chosen time step was 1.7 fs.

For a systematic study, first the melting points of nanoparticles were calculated via the potential energy and LI monitoring as a function of temperature. The nanoparticles were equilibrated during an initial (minimum) 250,000 time steps for each temperature, and then the potential energy and LI were calculated at each temperature during at least 150,000 steps depending on the system size. For some systems the equilibration needed more than 1000,000 time steps. The LI of each atom and of the entire system at each temperature are given as [21]:

$$\delta_i = 2\frac{1}{N-1}\sum_{j\neq i}^{N}\frac{\sqrt{<r_{ij}^2>_T - <r_{ij}>_T^2}}{<r_{ij}>_T}, \qquad (2)$$

$$\delta = \frac{1}{N}\sum_{i}^{N}\delta_i \qquad (3)$$

where $r_{ij}$ is distance between two atoms, N is number of atoms, $\delta_i$ is the LI of ith atom and $\delta$ is the LI of particle which is simply the normalized square root of mean distance among atoms. This parameter was widely used for solid-solid and solid-liquid transitions analysis and, as discussed in the previous works, the end of a sharp rise in the indices over a small temperature change reveals the melting point of system [22, 23]. Applying this index across the nanoparticles help analyzing the surface melting phenomenon. A reference Au bulk system (NVT) with a periodic boundary condition was simulated as well to compare against the current simulation results.

For studying the coalescence of nanoparticles, pairs of exactly the same equilibrated Au nanoparticles were brought to contact (separated by 0.2 nm) at different temperatures below and above the nanoparticle melting point and were studied for 200,000 time steps. No external forces were applied during the coalescence. The complete coalescence of each pair of the nanoparticles was defined for a temperature in which the diameter of the pair nanoparticle differed for less than 10% from its minimum value [24]. This definition is arbitrary but does not affect the conclusions drawn from the current studies.

## 3. Results and Discussion

Many theoretical and experimental studies show that the melting points of small particles are far lower than that of the corresponding bulk [25-28] that also has been confirmed by our simulations. Figures 1b, c illustrates the potential energy and LIs as a function of temperature for Au nanoparticles of different sizes. The jump in the values of these two quantities reveal the solid-liquid phase transition for which the corresponding temperature is the melting point. The results are shown in Table 1. A good agreement has been obtained between the current results and previous works [19, 28]. In order to obtain more accurate values of the melting temperatures, a detailed study of transition region is needed. As expected, the Gibbs-Thompson size effect on the melting point of nanoparticles is clearly obtained. Interestingly, the solid-liquid transformation energies (latent heats) are also found to be size-dependent, decreasing for smaller nanoparticles.

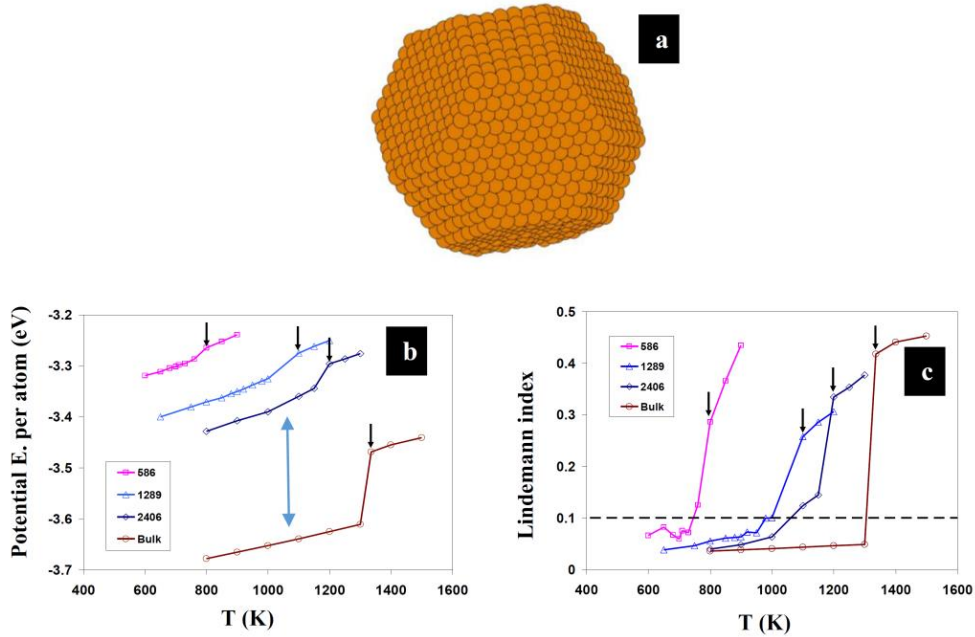

Figure 1: a) A typical Initial truncated polyhedral structure (Wolf Structure), b) Total potential energy per atom (eV) and c) LI versus temperature (K) during melting process of three Au nanoparticles with 586, 1289, 2406 atoms and Bulk Au material are shown.

The results in Figure 1b show that for a given temperature the potential energy levels for smaller systems are much higher than the bulk values (marked by the two-sided arrow). In fact, it is clear that this difference is much larger that the liquid-solid latent heat for the nanoparticles and almost comparable to the liquid-solid transformation energy in the bulk material. This implies the extended effect of the surfaces on the energy state of the nanoparticles, also discussed in our previous works, which is equivalent to the Laplace pressure induced by the surface energy. As a matter of fact, this extended surface effect is responsible for the decrease in the latent heat for smaller nanoparticles. As the size of particles increases the potential energy state approaches the bulk value.

Table 1. Results obtained from simulations. The differences between melting and coalescence temperatures increase for smaller nanoparticles. *This temperature is the lowest temperature in our simulation conditions that the coalescence was completed.

| number of atoms | diameter of particle (nm) ~ | melting point (K) ~ | approximated coalescence temperature* | $(T_m-T_c)/T_m$ |
|---|---|---|---|---|
| 586 | 2.8 | 800 | 600 | 0.25 |
| 1289 | 3.6 | 1100 | 900 | 0.18 |
| 2406 | 4.5 | 1200 | 1000 | 0.16 |
| Bulk | - | 1336 | - | - |

Figure 2a-c show the energy distributions across the nanoparticles for different temperatures (liquid and solid states). The surface-atoms (atoms at the surface or a couple of layers directly beneath it) are found to have higher energies. For the core of the particles, the potential energy is averaged and compared to the corresponding bulk value. Again on can see that the energy of the 'confined' bulk inside the nanoparticles is higher than the corresponding values in the nanoparticles (Laplace pressure). Moreover, it can be observed that a large number of surface-atoms in a solid-particle (for example $Au_{1289}$ at 750 K (figure 2b)) have the high energy values that are comparable to the values for the liquid state. This implies that a surface melting is possible even before reaching the melting temperature of the nanoparticle.

The LIs studies show the same upshots on the phase transition behavior of nanoparticles and bulk (figure 2d-f). The results indicate that the LIs are size-dependent and for a given temperature higher values for smaller systems (also please see figure 1c) are obtained. For the bulk sample, $\delta \geq 0.1$ corresponds to a melted phase but in an isolated nanoparticle the surface plays a destabilizing role.

As a result of the surface effect, the mean distance between atoms (mean-bond-length) and therefore the LI values are increased.

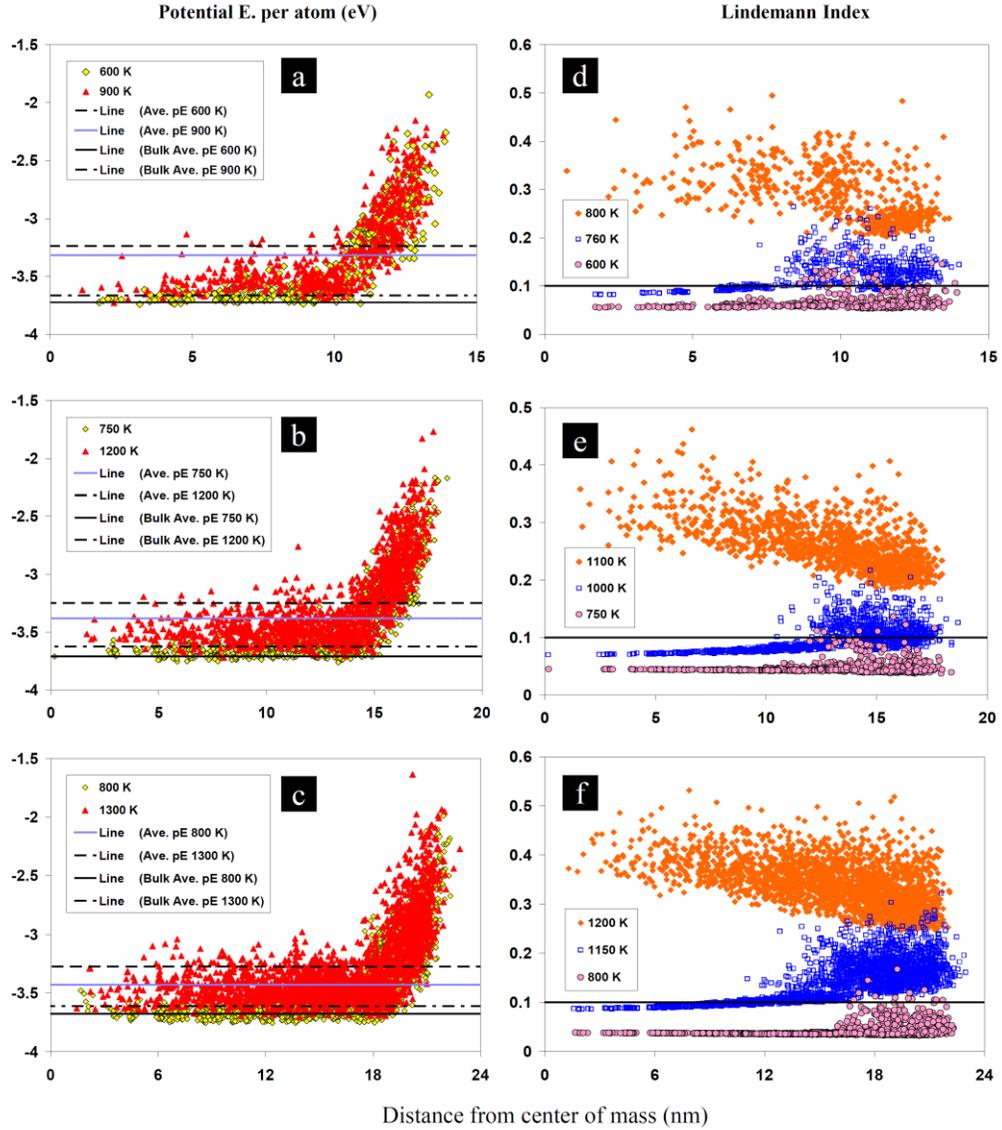

Figure 2: Potential energy distribution for solid and liquid states nanoparticles (potential energy of each atom (eV)), from center of mass to surface of a) $Au_{586}$, b) $Au_{1289}$, c) $Au_{2406}$. LI of each atom versus the distance of atom from center of mass for d) $Au_{586}$, e) $Au_{1289}$, f) $Au_{2406}$. The average potential energies for bulk and nanoparticles at the same temperatures are presented for comparison.

A larger scatter in the index values reflect the higher fluctuation and mobility of atoms on the surface of smaller particles. If we consider $\delta_i \geq 0.1$ as the threshold for melting (figure 2d-f), the LI values also indicate that surface melting is possible even though it is limited for the largest nanoparticle. Close to the melting temperature the mobility of atoms and the corresponding LIs rapidly increase.

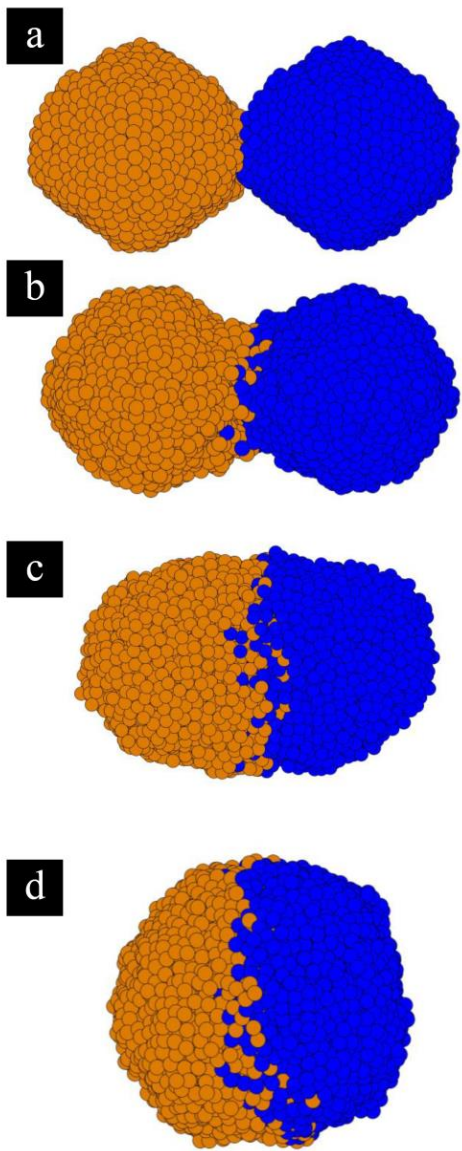

Figure 3: The snapshots of two coalescing $Au_{2406}$ nanoparticles at 1100 K, in four different time steps. Also see Figure 4.

The LIs studies confirm a surface premelting before reaching the melting point. This is clearly demonstrated in the LIs behavior Figure 2d-f. For smaller nanoparticles, surface melting occurs within a wider range of temperature. This critical phenomenon is important when the differences between coalescence and melting temperatures increase, by decrease the size of nanoparticles (See Table 1).

Figure 3 shows the evolution of two $Au_{2406}$ nanoparticles, initially at 1100 K. During the simulations, the initial contact between the two particles is established by the weak attractive forces between them. When two nanoparticles touch, the surface atoms quickly interact. A rapid diffusion of unstable surface-atoms from high-energy state (lower neighboring atoms) to the more stable low-energy state (more neighboring atoms) was observed. The corresponding size evaluations during the coalescence of the two nanoparticles in Figure 3 are marked in Figure 4a.

Figure 4a, b and c compare the evolution of two nanoparticles with different initial temperature, i.e. 1100 and 1250 K, in which the individual particles are in the solid and liquid states, respectively. The size evolution during the two simulations show a complete coalescence of the nanoparticles in both conditions. Analysis of the total potential energy and the average temperature of the two systems, however, reveal interesting details about the mechanisms of coalescence: For the liquid droplets (at 1250 K), coalescence occurs rapidly with a continuous increase in the temperature and potential energy state, that is due to the heat released upon the reduction in the total surface area. In case of the two solid nanoparticles at 1100 K a nontrivial path is taken in which an initial decrease (increase) in the total potential energy (temperature), is inversed sometime during the coalescence. The results show that in this case, the energy released due to the initial coalescence and surface reduction is invested in melting the nanoparticles that results in a drop in the average temperature and increase in the total potential

energy. In other words, a synergetic mechanism of coalescence is revealed in which the coalescence accelerates itself by the surface reduction.

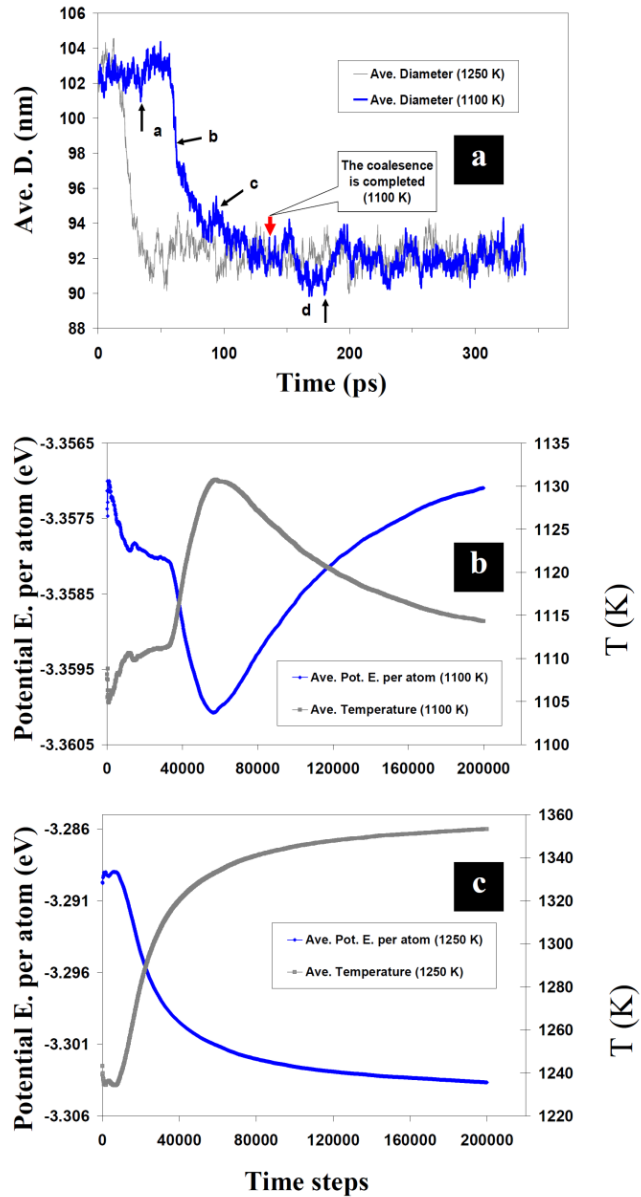

Figure 4: a) The size evaluations during coalescence of $Au_{2406}$ nanoparticles at 1100 (below the melting point) and 1250 K (above the melting point). The a-d points are the corresponding points for snapshots shown in Figure 3. The potential energy and temperature evaluations during coalescence of $Au_{2406}$ nanoparticles at b) 1100 and c) 1250 K.

The coalescence and reshaping of nanoparticle can have huge impact on their functional performance such as demonstrated in catalytic synthesis in chemical vapor deposition [29, 30]. Ding et al. have shown that the coalescence of iron clusters occurs at low temperatures without the clusters being molten [24]. They attributed this phenomenon to the curvature effect (Gibbs-Thompson effect) of iron clusters. The current results show that indeed solid state nanoparticles are also able to have a complete coalescence assisted by the surface diffusion mechanism. For small enough particles and/or at temperatures close enough to the melting point, a complete melting of the nanoparticles induced by the energy released due to the surface reduction is evidence in our studies. In fact, a large range of temperatures can be specified below melting point that liquid-solid phases may be coexisting in nanoparticles and a surface induced phase transition of the coalescing nanoparticles becomes possible. A detailed understanding of this phenomenon, however, needs to be more investigated especially from an atomic point of view.

Our results demonstrate that the leading mechanism of coalescence is surface diffusion (high energy and high mobility surface-atoms) followed by a rapid neck growth and a relatively slow reshaping of the attached nanoparticles. Our observations confirm the Foiles et al. [32] MD simulations results, however they did not account energy and temperature changes in the system. Our simulations show that at elevated temperature nanoparticles rapidly reshape while in lower temperatures (for example Au2406 at 800 K) the coalescence beginning quickly, but cannot be completed, because of the slow diffusion. Because of the surface area reduction, the coalescence is accompanied with an energy release increasing the temperature that accelerates the entire process (figure 4b and c). This confirms Lehtinen and Zachariah [31] reports on the sintering process of nanoparticles.

Although our simulations times were many orders of magnitude smaller than real experimental times, but the coalescences were completed for nanoparticles at several hundred degrees below their melting points. Therefore, in experimental times, it may be expected that lower coalescence temperatures obtain given a longer time for diffusion.

## 4. Conclusions

In current work, the melting and coalescence temperatures of Au nanoparticles and bulk were determined via MD simulation method. The potential energy and LI analyses were performed to understand the mechanism of melting and coalescence of nanoparticles.

The simulation results show that the coalescence occurs at temperatures lower than the corresponding melting points of the nanoparticles. With reference to LIs, it seems that the coalescence is accompanied with surface diffusion/melting. It was shown that the liquid-solid phases may be coexisting in a wide range of temperatures below the melting points in nanoparticles. Moreover, it was found that a surface-induced phase transition can occur where the heat released due to the reduction in the surface area can result in melting of the nanoparticles. This effect accelerates the coalescence process and can play a dominating role in small enough nanoparticles and/or at elevated temperature before the melting point.

**Acknowledgments**

The current manuscript is a translation of some parts of my master thesis on the *Thermodynamics of Nanosystems* (in Persian), 2006—2008, Iran. The software for performing the Molecular Dynamics Simulations and all analyses presented in this manuscript and in the thesis are developed by the author from scratch. Further details on this study is available in my thesis. The author would like to thanks the discussions with E. Marzban and the financial supports from the German Research Foundation, project DA 1655/2-1.